\documentclass[aps,prl,superscriptaddress,unsortedaddress,floatfix,showpacs,reprint]{revtex4-1}

\usepackage{verbatim}

\usepackage{amsmath}
\usepackage{amsfonts}
\usepackage{amssymb}
\usepackage{graphicx}
\usepackage{bm}
\usepackage{color}

\usepackage{epsf}

\usepackage{psfrag}

\begin{document}

\title
{Paving the Way Towards Precision Physics in Saturation Studies Through Exclusive Diffractive 
Light Neutral Vector Meson
Production 
}

\author{R.~Boussarie}
\affiliation{Institute of Nuclear Physics, Polish Academy of Sciences, Radzikowskiego 152, PL-31-342 Krak\'ow, Poland}

\author{A.~V.~Grabovsky}
\affiliation{Novosibirsk State University,
2 Pirogova street, Novosibirsk, Russia}
\affiliation{Budker Institute of Nuclear Physics,
11 Lavrenteva avenue, Novosibirsk, Russia}

\author{D.~Yu.~Ivanov}
\affiliation{Novosibirsk State University,
2 Pirogova street, Novosibirsk, Russia}
\affiliation{Sobolev Institute of Mathematics, 630090 Novosibirsk, Russia}

\author{L.~Szymanowski}
\affiliation{National Centre for Nuclear Research (NCBJ), Warsaw, Poland}

\author{S.~Wallon}
\affiliation{Laboratoire de Physique Th\'eorique (UMR 8627), CNRS, Univ. Paris-Sud,
Universit\'e Paris-Saclay, 91405 Orsay Cedex, France}
\affiliation{UPMC Univ. Paris 06, Facult\'e de Physique, 4 place Jussieu, 75252 Paris Cedex 05, France}

\begin{abstract}
\noindent
We perform the first next-to-leading order computation of the $\gamma^{(*)} \to V$ ($\rho, \phi, \omega$) impact factor
in the QCD shockwave approach and in the most general kinematics. This paves the way to the very first quantitative study of high-energy nucleon and nucleus saturation beyond the leading order, in various processes
to be measured in $ep$, $eA$, $pp$ and $pA$ collisions at existing and future colliders. 
\end{abstract}

\maketitle

\paragraph{Introduction.}

Among the various achievements of the HERA experiments, two landmark results emerged from $e^\pm p$ deep inelastic scattering (DIS).
First, 
diffractive events represent a fraction of up to 10\% of the total $e^\pm p$ cross-section for DIS~\cite{Ahmed:1995ns,*Adloff:1997sc,*Aktas:2006hx,*Aktas:2006hy,*Aaron:2010aa,*Aaron:2012ad,Derrick:1995wv,*Breitweg:1997aa,*Breitweg:1998gc,*Chekanov:2005vv,*Chekanov:2008fh,*Chekanov:2004hy,*Aaron:2012hua}. Second, the study of the kinematical domain where the photon virtuality $Q^2$ is moderate and the Bjorken $x$ variable is asymptotically small revealed that the proton saturates, both in inclusive and diffractive deep inelastic scattering, as first exhibited within the
Golec-Biernat and W\"usthoff  model~\cite{GolecBiernat:1998js,*GolecBiernat:1999qd}. 
It has been further realized that exclusive diffractive processes
could give an excellent lever arm to scrutinize the proton's internal structure at asymptotic energies. In particular, the exclusive diffractive production of a light vector meson $V$ ($\rho, \phi, \omega$)~\cite{Ivanov:2000uq,Munier:2001nr,Forshaw:2001pf,*Enberg:2003jw,*Poludniowski:2003yk} 
\begin{equation}
\label{process}
\gamma^{(*)} p \to V \, p
\end{equation}
was studied at HERA both for forward~\cite{Chekanov:2007zr,*Aaron:2009xp} and large $t$~\cite{Breitweg:1999jy,*Chekanov:2002rm,*Aktas:2006qs,*Aaron:2009xp} kinematics.
On top of the  photon virtuality, the transverse momenta
exchanged in the $t-$channel give access to the impact parameter distribution
of partons inside the proton {\it via} Fourier transformation.

Understanding the highly energetic proton state is  theoretically particularly appealing and phenomenologically valuable. First, at large center of mass energy $\sqrt{s}$, the proton is a dense system with high field strengths, still in the weak-coupling regime, and perturbative effective resummation methods must be applied. Second, in the context of relativistic heavy ion collisions, in view of producing and studying the quark-gluon plasma, the colliding nuclei in initial stages are saturated. Thus, saturation is one of the most important and longstanding problems
of QCD.
 
In different frameworks, 
either based on a QCD shockwave formalism~\cite{Balitsky:1995ub, *Balitsky:1998kc,*Balitsky:1998ya,*Balitsky:2001re}, a large-$N_c$
dipole model~\cite{Mueller:1993rr,*Mueller:1994jq,*Mueller:1994gb,*Chen:1995pa,Kovchegov:1999yj,*Kovchegov:1999ua} or
 an effective perturbative weak-coupling field theory
approach~\cite{JalilianMarian:1997jx,*JalilianMarian:1997gr,*JalilianMarian:1997dw,*JalilianMarian:1998cb,*Kovner:2000pt,*Weigert:2000gi,*Iancu:2000hn,*Iancu:2001ad,*Ferreiro:2001qy}, 
a color glass condensate (CGC) picture has emerged, describing the small-$x$ dynamics of QCD towards the saturation regime.

Still, it has been realized that in order to get a detailed understanding
of the properties of the high-energy proton, precise quantitative predictions are absolute prerequisites. This means that one should go beyond leading order computations, a task which is particularly difficult to achieve in the above mentioned frameworks. A first step towards such an improvement was performed at the level of the evolution kernel for CGC, including first the running coupling effects~\cite{Kovchegov:2006vj} and finally the whole next-to-leading order (NLO) correction to the kernel~\cite{Balitsky:2008zza,*Balitsky:2013fea,*Grabovsky:2013mba,*Balitsky:2014mca,Kovner:2013ona,*Lublinsky:2016meo,Caron-Huot:2015bja}. 
First steps have been made concerning the corrections to
the coupling to a probe. The fully inclusive NLO impact factor has been obtained 
for the coupling to a $\gamma^*$~\cite{Balitsky:2010ze,*Balitsky:2012bs}. The impact factor for semi-inclusive hadron production, involving the coupling to a parton, was computed in view of studying $p_\perp$-broadening effects~\cite{Chirilli:2011km,Ivanov:2012iv,Iancu:2016vyg}. Finally, the first computation of an exclusive NLO impact factor in the CGC framework was performed in Refs.~\cite{Boussarie:2014lxa,Boussarie:2016ogo}.

In this Letter, we study the exclusive production of a neutral longitudinally polarized vector meson with NLO accuracy.
As first noticed in Ref.~\cite{Brodsky:1994kf,*Frankfurt:1995jw}, one can
describe in DIS the exclusive production of a meson from a $q \bar{q}$ pair
based on
the collinear QCD factorization scheme. At moderate energies, the amplitude is given as  a convolution of quark or gluon generalized parton
distributions (GPDs) in the nucleon, the  distribution amplitude (DA) for the 
light meson,
and a perturbatively calculable
hard scattering amplitude~\cite{Collins:1996fb, Radyushkin:1997ki}. The DAs and GPDs are subject to specific QCD evolution equations~\cite{Farrar:1979aw,*Lepage:1979zb,*Efremov:1979qk}. Still, such a factorization is proven only for the twist-2 dominated transition between a longitudinally polarized photon and a longitudinally polarized vector meson~\cite{Collins:1996fb}. Explicit breaking of collinear factorization occurs at twist 3, through end-point singularities, in exclusive electroproduction of transversely polarized vector mesons~\cite{Mankiewicz:1999tt}. As a remedy, an improved collinear approximation scheme~\cite{Li:1992nu} has been proposed and
 applied to $\rho$ electroproduction~\cite{Vanderhaeghen:1999xj,*Goloskokov:2005sd,*Goloskokov:2006hr,*Goloskokov:2007nt}. At high energies, where the exchange of $t-$channel gluons dominates, $k_T-$factorization applies. The end-point singularities are naturally regularized by the transverse
momenta of these $t-$channel gluons~\cite{Ivanov:1998gk, Anikin:2009hk,*Anikin:2009bf}, providing models~\cite{Anikin:2011sa} to describe HERA data, including saturation effects~\cite{Besse:2012ia,*Besse:2013muy}.

In this Letter, we will carry out, for the first time in the shockwave context, a
complete NLO calculation for
exclusive diffractive meson production in $\gamma^{(*)} p$
or $\gamma^{(*)} A$ collisions, with completely general kinematics, by combining the collinear factorization and high energy small-$x$ factorization techniques. 
We will show the full infrared safe results for the $\gamma^\ast_L \rightarrow V_L$ and $\gamma^{(\ast)}_T \rightarrow V_L$ transitions. 
The details of the calculation will be provided in a separate article~\cite{Boussarie:2017}.

The present result provides the first calculation of higher order corrections,
in a complete NLO framework, of a vast class of processes.
Indeed, the complete generality of the kinematics allows it to be applied to a wide range of experimental conditions. 
It can describe either the electroproduction of vector mesons with general kinematics, or their photoproduction
at large transfered momentum. As a result,
it
can be used both at $ep$ and $eA$ colliders, like the future EIC~\cite{Boer:2011fh} or LHeC~\cite{AbelleiraFernandez:2012cc} and in ultraperipheral collisions  at RHIC or at the LHC~\cite{Baltz:2007kq,N.Cartiglia:2015gve}.

\paragraph{The shockwave framework.}

Let us define two lightlike vectors $n_{1}$ and $n_{2}$ such that
the partons in the upper (resp. lower) impact factor have large momentum
components along $n_{1}$ (resp. $n_{2}$). We write the Sudakov expansion of any vector $p$ as
\begin{eqnarray}
p^{\mu} & \equiv & p^{+}n_{1}^{\mu}+p^{-}n_{2}^{\mu}+p_{\perp}^{\mu}.\label{eq:lightconeVar}
\end{eqnarray}
Normalizing the lightcone basis so that $n_{1}\cdot n_{2}=1$, we write
the scalar product of two vectors as
\begin{eqnarray}
p\cdot q & \equiv & p^{+}q^{-}+p^{-}q^{+}+p_{\perp}\cdot q_{\perp} \nonumber \\ &=& p^{+}q^{-}+p^{-}q^{+}-\vec{p}\cdot\vec{q}\,.\label{eq:scalarProducts}
\end{eqnarray}
Within the shockwave formalism, the computation is performed in a frame where the target is highly boosted. We separate the gluonic field into an external (resp. internal) field containing
the gluons with momentum components along $n_{1}$ below (resp. above) the cutoff $e^\eta p_{\gamma}^{+}$, where $p_{\gamma}$ is the momentum of the photon and $\eta$ is the rapidity divide, which eventually separates the gluons belonging to the projectile impact factor from the ones attributed to the shockwave.

We work in the QCD lightcone gauge $n_2 \cdot A =0.$
In the high energy limit and in this gauge, the external field $b_{\eta}^{\mu}$ is located at zero lightcone time $z^+$ and has the eikonal Lorentz structure  
\begin{equation}
b_{\eta}^{\mu}(z)=b_{\eta}^{-}(\vec{z}\,)\, \delta(z^{+})\, n_{2}^{\mu}.\label{eq:ExternalField}
\end{equation}
We define the high-energy Wilson line operator as
\begin{equation}
U_{\vec{z}}^{\eta} \equiv \mathcal{P}\exp\left[ig \! \int_{-\infty}^{+\infty} \! dz^{+}b_{\eta}^{-}(z)\right].\label{eq:WilsonLine}
\end{equation}
The scattering amplitude is obtained by convoluting the impact factor
with the matrix elements of
operators built from Wilson line operators acting on the target states.

In the context of such a NLO diffractive process, we introduce the dipole and double dipole operators in momentum space
from  Wilson line operators 
in the
fundamental representation of $SU(N_c)$
as
\begin{eqnarray}
& &\left[\mathrm{Tr}(U_{1}^{\eta}U_{2}^{\eta\dagger})-N_{c}\right](\vec{p}_{1},\,\vec{p}_{2})\label{eq:DipoleOperator}\\
&\equiv&\int \! d^{d}\vec{z}_{1}d^{d}\vec{z}_{2} \, e^{-i(\vec{p}_{1}\cdot\vec{z}_{1})-i(\vec{p}_{2}\cdot\vec{z}_{2})}\left[\mathrm{Tr}(U_{\vec{z}_{1}}^{\eta}U_{\vec{z}_{2}}^{\eta\dagger})-N_{c}\right],\nonumber 
\end{eqnarray}
and
\begin{eqnarray}
& & \left[\mathrm{Tr}(U_{1}^{\eta}U_{3}^{\eta\dagger})\mathrm{Tr}(U_{3}^{\eta}U_{2}^{\eta\dagger})-N_{c}\mathrm{Tr}(U_{1}^{\eta}U_{2}^{\eta\dagger})\right](\vec{p}_{1},\,\vec{p}_{2},\,\vec{p}_{3})\nonumber\\
&\equiv&\int \! d^{d}\vec{z}_{1}d^{d}\vec{z}_{2}d^d\vec{z}_3 \, e^{-i(\vec{p}_{1}\cdot\vec{z}_{1})-i(\vec{p}_{2}\cdot\vec{z}_{2})-i(\vec{p}_{3}\cdot\vec{z}_{3})} \label{eq:DoubleDipoleOperator} \\
&\times&\left[\mathrm{Tr}(U_{\vec{z}_{1}}^{\eta}U_{\vec{z}_{3}}^{\eta\dagger})\mathrm{Tr}(U_{\vec{z}_{3}}^{\eta}U_{\vec{z}_{2}}^{\eta\dagger})-N_{c}\mathrm{Tr}(U_{\vec{z}_{1}}^{\eta}U_{\vec{z}_{2}}^{\eta\dagger})\right],\nonumber 
\end{eqnarray}
where $d=2+2 \epsilon$ is the transverse dimension.
In these equations, $\vec{z}_1,\vec{z}_2,\vec{z}_3$ are respectively the transverse coordinates of the interaction points of the quark, the antiquark and the gluon with the external shockwave field. Their conjugate transverse momenta  $\vec{p}_1,\vec{p}_2,\vec{p}_3$ are the incoming effective momenta acquired {\it via} interaction with the $t-$channel shockwave field.

\paragraph{Factorization scheme.}

\begin{figure}
\hspace{-0cm}\raisebox{0cm}{\rotatebox{0}{\psfrag{D}{1}
\includegraphics[scale=0.21]{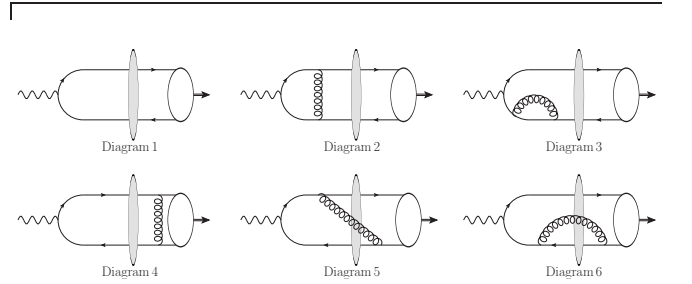}}}
\caption{Contributions to the impact factor for $\gamma^* \to V$ transition. The gray blobs stand for the external (shockwave) field while the white blobs denote the distribution amplitudes of the produced vector meson.}
\label{fig:diagrams}
\end{figure}

At leading order accuracy, the factorized amplitude is the action
of an operator $\mathcal{A}_{LO}^\eta$ on target states. This operator
is the convolution of the dipole operator, a hard part $\Phi_0$ to which we will refer as the impact factor, and a DA. The twist 2 DA $\varphi$
for a longitudinally polarized vector meson $V_L$ is
 defined {\it via} the matrix element of a non-local lightcone operator renormalized at scale $\mu_F$
\begin{eqnarray}
\label{def:DA}
&&\langle V_L(p_V)|\bar \Psi(y) \gamma^\mu \Psi(0)|0 \rangle_{y^2 \to 0}  \nonumber \\
&&
 = f_V\, p_V^\mu \!
\int^1_0 \!dx\, e^{ix(p_V \cdot y)}\, \varphi (x, \mu_F)\, ,
\end{eqnarray}
where the gauge link between fields was omitted since it does not contribute in the chosen lightcone gauge. We write the operator as follows:
\begin{eqnarray}
\mathcal{A}_{LO}^\eta &\equiv  &-\frac{e_{V} \, f_V \, \varepsilon_\beta}{N_{c}}\int_{0}^{1}\!\!dx \, \varphi\left(x,\mu_F\right)\int \!\frac{d^{d}\vec{p}_{1}}{\left(2\pi\right)^{d}}\frac{d^{d}\vec{p}_{2}}{\left(2\pi\right)^{d}} \label{eq:LOfac}
 \\
& \times & \left(2\pi\right)^{d+1} \delta\left(p_{V}^{+}-p_{\gamma}^{+}\right)\delta\left(\vec{p}_{V}-\vec{p}_{\gamma}-\vec{p}_{1}-\vec{p}_{2}\right)\nonumber \\
 & \times & \Phi_{0}^\beta\left(x,\,\vec{p}_{1},\,\vec{p}_{2}\right)\left[\mathrm{Tr}(U_{1}^\eta U_{2}^{\eta\dagger})-N_{c}\right]\left(\vec{p}_{1},\,\vec{p}_{2}\right).
 \nonumber
\end{eqnarray}
Here $\varepsilon_\beta$ is the polarization vector of the photon, $f_V$ is the meson coupling which is related to the vector meson decay into leptons and $e_V$ is an effective electric quark charge which takes into account the flavor content of the meson~\footnote{For example for a $\rho_0$ it reads $e_V = \frac{e_u - e_d}{\sqrt{2}} = \frac{e}{\sqrt{2}}$.}. 
$\Phi_0$ is obtained by computing diagram 1 in Fig.~\ref{fig:diagrams} using the effective shockwave Feynman rules~\cite{Boussarie:2016ogo}. 

In practice to obtain a full physical amplitude one should first solve the Jalilian-Marian-Iancu-McLerran-Weigert-Leonidov-Kovner (JIMWLK)~\cite{JalilianMarian:1997jx,*JalilianMarian:1997gr,*JalilianMarian:1997dw,*JalilianMarian:1998cb,*Kovner:2000pt,*Weigert:2000gi,*Iancu:2000hn,*Iancu:2001ad,*Ferreiro:2001qy} evolution equation for the Wilson line operators, which here reduces to the dipole Balitsky-Kovchegov (BK)~\cite{Balitsky:1995ub, *Balitsky:1998kc,*Balitsky:1998ya,*Balitsky:2001re,Kovchegov:1999yj,*Kovchegov:1999ua} evolution, and act on the target states. 
For example in the case of a scattering off a proton, the leading order amplitude
$A_{LO}^\eta\equiv\left\langle P'|\mathcal{A}^\eta_{LO}|P\right\rangle $
will be given in terms of the non-forward dipole-proton scattering amplitude 
\begin{equation}
\label{DipoleAmplitude}
\left\langle P'\right|[\mathrm{Tr}(U_{1}^\eta U_{2}^{\eta\dagger})-N_{c}]\left(\vec{p}_{1},\,\vec{p}_{2}\right)\left|P\right\rangle \,.
\end{equation}

At NLO accuracy the double dipole operator starts to contribute and we define similarly to the LO case the NLO operator
\begin{widetext}\begin{eqnarray}
\label{eq:NLOfac}
\mathcal{A}_{NLO}^\eta & \equiv & -\frac{e_{V} f_V \varepsilon_\beta}{N_{c}}\int_{0}^{1}\!dx \, \varphi\left(x,\mu_F\right)\int\!\frac{d^{d}\vec{p}_{1}}{\left(2\pi\right)^{d}}\frac{d^{d}\vec{p}_{2}}{\left(2\pi\right)^{d}}\frac{d^{d}\vec{p}_{3}}{\left(2\pi\right)^{d}}\left(2\pi\right)^{d+1}\delta\left(p_{V}^{+}-p_{\gamma}^{+}\right)\delta\left(\vec{p}_{V}-\vec{p}_{\gamma}-\vec{p}_{1}-\vec{p}_{2}-\vec{p}_{3}\right)\nonumber \\
 &  & \times\frac{\alpha_{s}\Gamma\left(1-\epsilon\right)}{\left(4\pi\right)^{1+\epsilon}}\left\{ \left(\frac{N_{c}^{2}-1}{N_{c}}\right)\Phi_{1}^{\beta}(x,\,\vec{p}_{1},\,\vec{p}_{2})\left[\mathrm{Tr}(U_{1}^\eta U_{2}^{\eta\dagger})-N_{c}\right]\left(\vec{p}_{1},\,\vec{p}_{2}\right)\left(2\pi\right)^{d}\delta\left(\vec{p}_{3}\right)\right.\nonumber \\
 &  & \left.+ \, \Phi_{2}^{\beta}\left(x,\,\vec{p}_{1},\,\vec{p}_{2},\,\vec{p}_{3}\right)\left[\mathrm{Tr}(U_{1}^{\eta\dagger} U_{3}^{\eta\dagger})\mathrm{Tr}(U_{3}^{\eta\dagger} U_{2}^{\eta\dagger})-N_{c}\mathrm{Tr}(U_{1}^\eta U_{2}^{\eta\dagger})\right]\left(\vec{p}_{1},\,\vec{p}_{2},\,\vec{p}_{3}\right)\right\}.
\end{eqnarray}\end{widetext}
The explicit expressions for $\Phi_1$ and $\Phi_2$, given below in Eqs.~(\ref{eq:ResultNLO1L}-\ref{eq:Phi2T}), 
are the main results of the present Letter.
Again, in the example of the scattering on a proton, the computation of the NLO amplitude 
$A_{NLO}^\eta\equiv\left\langle P'|\mathcal{A}^\eta_{NLO}|P\right\rangle $
will now involve, in addition to the amplitude~(\ref{DipoleAmplitude}), the non-forward double-dipole-proton scattering amplitude 
\begin{eqnarray}
\label{DoubleDipoleAmplitude}
&&\langle P' |
[\mathrm{Tr}(U_{1}^{\eta\dagger} U_{3}^{\eta\dagger})
\mathrm{Tr}(U_{3}^{\eta\dagger} U_{2}^{\eta\dagger}) 
\nonumber \\
&&-N_{c}\mathrm{Tr}(U_{1}^\eta U_{2}^{\eta\dagger})]\left(\vec{p}_{1},\,\vec{p}_{2},\,\vec{p}_{3}\right)
| P \rangle\,.
\end{eqnarray}
In order to get phenomenological predictions for the whole process~(\ref{process}) at NLO, one should combine the NLO impact factors $\Phi_1$ and $\Phi_2$  with 
the two scattering amplitudes~(\ref{DipoleAmplitude}, \ref{DoubleDipoleAmplitude}),
which are
obtained by solving the NLO JIMWLK equation 
with initial conditions at rapidity $\eta_0$ with
$p^+_{\rm target} = e^{\eta_0} p_\gamma^+$. Then
\begin{equation}
\eta - \eta_0 = \ln \frac{s}{s_0} \,,
\end{equation}
where the arbitrary scale $s_0 \sim \, p_{\rm target}^+ \, p_{\rm target}^- \ll s$ is a typical target scale.

In Eq.~(\ref{eq:NLOfac}),
$\Phi_2$ is obtained by computing diagrams 5 and 6 (and their $q \leftrightarrow \bar{q}$ symmetric counterparts) with $\vec{p}_3\neq\vec{0}$, see Fig.~\ref{fig:diagrams}. $\Phi_1$ is the sum of the $\vec{p}_3=\vec{0}$ contribution from the same two diagrams with the contribution from diagrams 2, 3 (and its $q \leftrightarrow \bar{q}$ symmetric counterpart) and 4.
For readability we will now omit the dependence on the $t$-channel transverse momenta in the impact factors $\Phi_i$.
The QED gauge invariance relation
\begin{equation}
p_\gamma \cdot \Phi_i = 0
\end{equation}
for $i=0,1,2$ allows one to reduce the computation to the only evaluation of $\Phi_i^+$ and $\Phi_{i\perp}^\beta$. In the following we will work in the frame where the transverse momentum of the photon is $\vec{0}$. The contributions to the $\gamma^*_L \to V_L$ and $\gamma^{(*)}_T \to V_L$ transitions are then given by
\begin{equation}
\label{transitionLT}
\varepsilon_L \cdot \Phi_i = \frac{Q}{p_\gamma^+} \Phi_i^+ \quad {\rm and} \quad \varepsilon_T \cdot \Phi_i = \varepsilon_\perp \cdot \Phi_{i\perp}\,.
\end{equation}

\paragraph{Divergences and evolution equations.}

First, let us note that in the shockwave framework, contrary to the Balitsky-Fadin-Kuraev-Lipatov (BFKL)~\cite{Fadin:1975cb,* Kuraev:1976ge,*Kuraev:1977fs,*Balitsky:1978ic} approach, the coupling to the $t-$channel
exchanged state does not involve the QCD coupling constant. As a consequence, the LO impact factor as defined in Eq.~(\ref{eq:LOfac}) is of order $\alpha_s^0$, while the NLO impact factor in Eq.~(\ref{eq:NLOfac}) is of order $\alpha_s.$
Thus the running of $\alpha_s$ has to be considered as an NNLO effect
when computing an impact factor.

The intermediate steps of the calculation involve various types
of divergences, namely ultraviolet, soft, collinear and the spurious lightcone gauge pole (to which we will refer as the rapidity divergence).
These divergences are controlled by dimensional regularization in transverse
space $d\equiv2+2\epsilon$, and by an infinitesimal
cutoff $\alpha p_\gamma^+$ in longitudinal space. In particular the rapidity
divergence, which is regularized by the $\alpha$ cut-off, is canceled {\it via} the BK-JIMWLK evolution equation for the
dipole operator, which allows one to get rid of the dependence
on $\alpha$. In momentum space it reads~\cite{Boussarie:2016ogo}
\begin{widetext}
\begin{eqnarray}
\label{eq:BKJIMWLK}
& & \frac{\partial}{\partial\eta} \left[\mathrm{Tr}(U_{1}^{\eta}U_{2}^{\eta\dagger})-N_{c}\right](\vec{p}_{1},\,\vec{p}_{2}) = \alpha_{s}\mu^{2-d}\int\frac{d^{d}\vec{k}_{1}d^{d}\vec{k}_{2}d^{d}\vec{k}_{3}}{(2\pi)^{2d}}\delta(\vec{k}_{1}+\vec{k}_{2}+\vec{k}_{3}-\vec{p}_{1}-\vec{p}_{2}) \\
& \times & \mathcal{H}(\vec{k}_1,\vec{k}_2,\vec{k}_3,\vec{p}_1,\vec{p}_2) \left[\mathrm{Tr}(U_{1}^{\eta}U_{3}^{\eta\dagger})\mathrm{Tr}(U_{3}^{\eta}U_{2}^{\eta\dagger})-N_{c}\mathrm{Tr}(U_{1}^{\eta}U_{2}^{\eta\dagger})\right](\vec{k}_{1},\,\vec{k}_{2},\,\vec{k}_{3}),\nonumber  
\end{eqnarray}
\end{widetext}
where the kernel $\mathcal{H}$ reads
\begin{eqnarray}
& & \mathcal{H}(\vec{p}_1,\vec{p}_2,\vec{p}_3,\vec{k}_1,\vec{k}_2)=4\frac{(\vec{k}_{1}-\vec{p}_{1})\cdot(\vec{k}_{2}-\vec{p}_{2})}{(\vec{k}_{1}-\vec{p}_{1})^{2}(\vec{k}_{2}-\vec{p}_{2})^{2}} \label{BKkernel} \\ &&+ \frac{\Gamma(1-\frac{d}{2})\Gamma^{2}(\frac{d}{2})}{\Gamma(d-1)}\!\left(\!\frac{2\pi^{\frac{d}{2}}\delta(\vec{k}_{2}-\vec{p}_{2})}{\left[(\vec{k}_{1}-\vec{p}_{1})^{2}\right]^{1-\frac{d}{2}}}+\frac{2\pi^{\frac{d}{2}}\delta(\vec{k}_{1}-\vec{p}_{1})}{\left[(\vec{k}_{2}-\vec{p}_{2})^{2}\right]^{1-\frac{d}{2}}}\!\right)\!. \nonumber
\end{eqnarray}
Evolving the Wilson lines from $\alpha$ to $\eta$ creates a counterterm to the double dipole contribution as follows:
\begin{eqnarray}
\label{JIMWLKcounterterm}
&  &  
\tilde{\Phi}_{2}^{\beta}(\eta,\alpha,\vec{p}_{1},\vec{p}_{2},\vec{p}_{3}) =  -\frac{\mu^{2-d}}{\Gamma(1-\epsilon)\pi^{1+\epsilon}}\ln\left(\frac{e^{\eta}}{\alpha}\right) \nonumber \\ & & 
\times 
\! \int \! d^{d}\vec{k}_{1}d^{d}\vec{k}_{2}\,\delta(\vec{p}_{V}-\vec{p}_{\gamma}-\vec{k}_{1}-\vec{k}_{2}) 
\\ &  & 
\times
\mathcal{H}(\vec{p}_{1},\,\vec{p}_{2},\,\vec{p}_{3},\,\vec{k}_{1},\,\vec{k}_{2})\,\Phi_{0}^{\beta}(x,\,\vec{k}_{1},\,\vec{k}_{2}) \,
\nonumber
.\hspace{.5cm}
\end{eqnarray}
This counterterm allows one to get rid of the dependence on $\alpha$ in the NLO contribution. By similar arguments one can cancel the overall dependence of the impact factor on the rapidity divide $\eta$ up to NNLO terms: indeed 
changing $\eta$ to $\eta'$ and
evolving the LO amplitude from $\eta$ to $\eta^\prime$ gives rise to $\tilde{\Phi}_{2}^{\beta}(\eta^\prime,\eta,\vec{p}_{1},\vec{p}_{2},\vec{p}_{3})$, whose dependence on $\eta$ is canceled by combining it with the NLO impact factor.

The collinear divergence is cancelled via the Efremov-Radyushkin-Brodsky-Lepage (ERBL) evolution equation
for the twist 2 DA $\varphi$. In the $\overline{MS}$ scheme it reads 
\begin{eqnarray}
\label{eq:ERBLevolution}
\frac{\partial \varphi(x,\mu_F)}{\partial\ln\mu^{2}_F} 
=\frac{\alpha_{s}C_{F}}{2\pi}\frac{\Gamma(1-\epsilon)}{(4\pi)^{\epsilon}}\!\left(\!\frac{\mu^2_F}{\mu^2}\!\right)^\epsilon
\!\!\! 
\int_{0}^{1}\!\!\!dz\, \varphi(z,\mu_F)\,\mathcal{K}(x,z),
\nonumber
\\
\end{eqnarray}
where $C_F\equiv(N_c^2-1)/(2 N_c)$ is the Casimir in the fundamental
representation of $SU(N_c)$ and
 $\mathcal{K}(x,z)$ is the well known ERBL evolution kernel~\cite{Farrar:1979aw,Lepage:1979zb,Efremov:1979qk}
\begin{widetext}
\begin{eqnarray}
\label{eq:ERBLkernel}
\mathcal{K}(x,z) = \frac{1-x}{1-z}\left(1+\left[\frac{1}{x-z}\right]_+\right)\theta(x-z)  
+\frac{x}{z}\left(1+\left[\frac{1}{z-x}\right]_+\right)\theta(z-x)+ \frac{3}{2} \,\delta(z-x) . 
\end{eqnarray}
\end{widetext}
For a function $F(z)$ which behaves as $F_0+F_1 \ln(z-z_0)$ for $z\rightarrow z_0$ we defined the $+$ prescription as
\begin{equation}
\int_0^1 \!\! dz \left[\frac{1}{z-z_0}\right]_+ \! F(z) \equiv \int_0^1 \!\! dz \frac{F(z)-F_0-F_1 \ln(z-z_0)}{z-z_0} \label{eq:PlusPrescription}.
\end{equation}
Evolving the DA in the LO contribution from 0 to $\mu_F$ gives rise to a counterterm to the NLO dipole contribution, which reads
\begin{equation}
\tilde{\Phi}_{1}^{\beta}(x,\mu_{F}) = -\!\int_{0}^{1}\!\!\! dz\,\mathcal{K}(z,x)\left[\frac{1}{\epsilon}+\ln\left(\frac{\mu_{F}^{2}}{\mu^{2}}\right)\right]\!\Phi_{0}^{\beta}(z). \label{ERBLcounterterm}
\end{equation}

\paragraph{Infrared finiteness and final results.}

The leading order impact factor reads
\begin{eqnarray}
\label{eq:LOif}
\Phi_{0}^{+}(x) & = & \frac{2x\bar{x}\left(p_{V}^{+}\right)^{2}}{\left[\left(\bar{x}\vec{p}_{1}-x\vec{p}_{2}\right)^{2}+x\bar{x}Q^{2}\right]}\,,
\\
\label{eq:LOifTrans}
\Phi_{0\perp}^{\beta}(x) & = & \frac{(x-\bar{x})p_{V}^{+}(\bar{x}p_{1\perp}^{\beta}-xp_{2\perp}^{\beta})}{\left[\left(\bar{x}\vec{p}_{1}-x\vec{p}_{2}\right)^{2}+x\bar{x}Q^{2}\right]} \, ,
\end{eqnarray}
where $\bar{x} \equiv 1-x.$

Let us consider separately the NLO dipole contribution $\Phi_1$ and the double dipole contribution $\Phi_2$ since they are independently gauge invariant and infrared finite, and the mechanisms for the cancellation of their divergences are different.

The sum of the dipole contribution from each diagram with the contribution in Eq.~(\ref{ERBLcounterterm}) from the ERBL evolution of the DA
is finite. 
It reads:
\begin{widetext} 
\begin{eqnarray}
\label{eq:ResultNLO1L}
\Phi_{1}^{+}\left(x\right) & = & \int_{0}^{x}dz\left(\frac{x-z}{x}\right)\left[1+\left(1+\left[\frac{1}{z}\right]_{+}\right)\ln\left(\frac{\left(\left((\bar{x}+z)\vec{p}_{1}-(x-z)\vec{p}_{2}\right)^{2}+(x-z)(\bar{x}+z)Q^{2}\right)^{2}}{\mu_{F}^{2}(x-z)(\bar{x}+z)Q^{2}}\right)\right]\Phi_{0}^{+}\left(x-z\right)\nonumber \\
 & + & \frac{1}{2}\Phi_{0}^{+}\left(x\right)\left[\frac{1}{2}\ln^{2}\left(\frac{\bar{x}}{x}\right)+3-\frac{\pi^{2}}{6}-\frac{3}{2}\ln\left(\frac{\left((\bar{x}\vec{p}_1 -x\vec{p}_2)^{2}+x\bar{x}Q^{2}\right)^{2}}{x\bar{x}\mu_{F}^{2}Q^{2}}\right)\right] 
 \nonumber \\ 
 & + & \frac{\left(p_{\gamma}^{+}\right)^{2}}{2x\bar{x}}\int_{0}^{x}dz\left[\left(\phi_{5}\right)_{LL}|_{\vec{p}_{3}=\vec{0}}+\left(\phi_{6}\right)_{LL}|_{\vec{p}_{3}=\vec{0}}\right]_{+}+(x\leftrightarrow\bar{x},\vec{p}_1 \leftrightarrow \vec{p}_2) 
\end{eqnarray}
\end{widetext}
for a longitudinal photon, and
\begin{widetext}
\begin{eqnarray}
\label{eq:resultNLOdipolePerp-1}
\Phi_{1\perp}^{\beta}\left(x\right) & = & \frac{1}{4}\left[\ln^{2}\left(\frac{\bar{x}}{x}\right)-\frac{\pi^{2}}{3}+6-3\ln\left(\frac{(\bar{x}\vec{p}_{1}-x\vec{p}_{2})^{2}+x\bar{x}Q^{2}}{\mu_{F}^{2}}\right)
 +3\frac{x\bar{x}Q^{2}}{(\bar{x}\vec{p}_{1}-x\vec{p}_{2})^{2}}\ln\left(\frac{(\bar{x}\vec{p}_{1}-x\vec{p}_{2})^{2}+x\bar{x}Q^{2}}{x\bar{x}Q^{2}}\right)\right]\Phi_{0\perp}^{\beta}\left(x\right)
 \nonumber \\
 &  & +\int_{0}^{x}dz\left(\frac{x-z}{x}\right)\Phi_{0\perp}^{\beta}\left(x-z\right)\left[1+\left(1+\left[\frac{1}{z}\right]_{+}\right)\ln\left(\frac{(\left(\bar{x}+z\right)\vec{p}_{1}-\left(x-z\right)\vec{p}_{2})^{2}+\left(x-z\right)\left(\bar{x}+z\right)Q^{2}}{\mu_{F}^{2}}\right)\right.\nonumber \\
 &  & \left.-\left(1+\left[\frac{1}{z}\right]_{+}\right)\frac{\left(x-z\right)\left(\bar{x}+z\right)Q^{2}}{(\left(\bar{x}+z\right)\vec{p}_{1}-\left(x-z\right)\vec{p}_{2})^{2}}\ln\left(\frac{(\left(\bar{x}+z\right)\vec{p}_{1}-\left(x-z\right)\vec{p}_{2})^{2}+\left(x-z\right)\left(\bar{x}+z\right)Q^{2}}{\left(x-z\right)\left(\bar{x}+z\right)Q^{2}}\right)\right]\nonumber \\
 &  & +\frac{p_{\gamma}^{+}}{2x\bar{x}}\int_{0}^{x}dz\left[\left(\phi_{5}\right)_{TL}^\beta|_{\vec{p}_{3}=\vec{0}}+\left(\phi_{6}\right)_{TL}^\beta|_{\vec{p}_{3}=\vec{0}}\right]_{+}+\left(x\leftrightarrow\bar{x},\vec{p}_{1}\leftrightarrow\vec{p}_{2}\right) 
\end{eqnarray}
\end{widetext}
for a transverse photon. The quantities $(\phi_{5,6})_{LL}$ and $(\phi_{5,6})_{TL}^\beta$ can be extracted from Ref.~\cite{Boussarie:2016ogo}, with the change of variables $(\vec{p}_q,\vec{p}_{\bar{q}}) \rightarrow (x \vec{p}_V, \bar{x}\vec{p}_V)$~\footnote{see Eqs.~(A.22), (A.24), (A.27), (A.29) in Ref.~\cite{Boussarie:2016ogo}}. One should understand $[\phi]_+$ in Eqs.~(\ref{eq:ResultNLO1L}) and (\ref{eq:resultNLOdipolePerp-1}) as the finite term which results from the replacement of the $\frac{1}{z}$ pole in $\phi$ by the $+$ prescription as defined in Eq.~(\ref{eq:PlusPrescription}).
The total dependence on the dimensional regulator $\mu$ cancels as expected, and the absence of a renormalization scale is due to the absence of running coupling contributions in the impact factor at this order. In the final expressions, $\phi_{6}$ terms are evaluated by replacing  $\mu$ by $\mu_F$ in Ref.~\cite{Boussarie:2016ogo}.

The sum of the double dipole contributions with the contribution~(\ref{JIMWLKcounterterm}) from the BK-JIMWLK evolution of the dipole operator is finite and reads
\begin{widetext}
\begin{eqnarray}
\Phi_{2}^{+} & = & -\frac{x\bar{x}\left(p_{\gamma}^{+}\right)^{2}\left(\left(x\vec{p}_{V}-\vec{p}_{1}\right)^{2}+\left(\bar{x}\vec{p}_{V}-\vec{p}_{2}\right)^{2}-\vec{p}_{3}^{\,\,2}+2x\bar{x}Q^{2}\right)}{\left(\left(x\vec{p}_{V}-\vec{p}_{1}\right)^{2}+x\bar{x}Q^{2}\right)\left(\left(\bar{x}\vec{p}_{V}-\vec{p}_{2}\right)^{2}+x\bar{x}Q^{2}\right)-x\bar{x}\vec{p}_{3}^{\,\,2}Q^{2}}\nonumber \\
 &  & \times\ln\left(\frac{x\bar{x}}{e^{2\eta}}\right)\ln\left(\frac{\left(\left(x\vec{p}_{V}-\vec{p}_{1}\right)^{2}+x\bar{x}Q^{2}\right)\left(\left(\bar{x}\vec{p}_{V}-\vec{p}_{2}\right)^{2}+x\bar{x}Q^{2}\right)}{x\bar{x}\vec{p}_{3}^{\,\,2}Q^{2}}\right)\label{eq:Phi2L} \\
 & - & \frac{4x\bar{x}\left( p_{\gamma}^{+}\right)^2}{\left(x\vec{p}_{V}-\vec{p}_{1}\right)^{2}+x\bar{x}Q^{2}}\ln\left(\frac{\bar{x}}{e^{\eta}}\right)\ln\left(\frac{\vec{p}_{3}^{\,\,2}}{Q^2}\right)+\frac{\left(p_{\gamma}^{+}\right)^{2}}{2x\bar{x}}\int_{0}^{x}dz\left[\left(\phi_{5}\right)_{LL}+\left(\phi_{6}\right)_{LL}\right]_{+}+(x\leftrightarrow\bar{x},\vec{p}_1 \leftrightarrow \vec{p}_2) \nonumber
\end{eqnarray}
for a longitudinal photon, and 
\begin{eqnarray}
\label{eq:Phi2T}
&&\Phi_{2\perp}^{\beta}\left(x\right)  =  p_{\gamma}^{+}(xp_{V\perp}^{\beta}-p_{1\perp}^{\beta})\left(\bar{x}-x\right)\left(\frac{-2}{(x\vec{p}_{V}-\vec{p}_{1})^{2}+x\bar{x}Q^{2}}\ln\left(\frac{\vec{p}_{3}^{\,\,2}}{Q^2}\right)\ln\left(\frac{\bar{x}}{e^{\eta}}\right)\right. \\
 &  & +\ln\left(\frac{x\bar{x}}{e^{2\eta}}\right)\!\!\left[\frac{1}{(x\vec{p}_{V}-\vec{p}_{1})^{2}}\ln\left(\!\frac{(x\vec{p}_{V}-\vec{p}_{1})^{2}+x\bar{x}Q^{2}}{x\bar{x}Q^{2}}\!\right)
 \!-\!\left.\frac{\left((\bar{x}\vec{p}_{V}-\vec{p}_{2})^{2}+x\bar{x}Q^{2}\right)\ln\left(\!\frac{\left((x\vec{p}_{V}-\vec{p}_{1})^{2}+x\bar{x}Q^{2}\right)\left((\bar{x}\vec{p}_{V}-\vec{p}_{2})^{2}+x\bar{x}Q^{2}\right)}{x\bar{x}\vec{p}_{3}^{\,\,2}Q^{2}}\!\right)}{\left((x\vec{p}_{V}-\vec{p}_{1})^{2}+x\bar{x}Q^{2}\right)\left((\bar{x}\vec{p}_{V}-\vec{p}_{2})^{2}+x\bar{x}Q^{2}\right)-x\bar{x}\vec{p}_{3}^{\,\,2}Q^{2}}\!\right]\!\right)\!\!
 \nonumber \\ &  & 
 +\frac{p_{\gamma}^{+}}{2x\bar{x}}\int_{0}^{x}dz\left[\left(\phi_{5}\right)_{TL}^{\beta}+\left(\phi_{6}\right)_{TL}^{\beta}\right]_++\left(x\leftrightarrow\bar{x},\vec{p}_1\leftrightarrow\vec{p}_2\right)\nonumber 
\end{eqnarray}
\end{widetext}
for a transverse photon.

\paragraph{Discussion.}

An explicit check shows that Eqs.~(\ref{eq:resultNLOdipolePerp-1}, \ref{eq:Phi2T}) are still valid in the $Q^2=0$ (photoproduction) limit despite the presence of $\ln Q^2$ terms since they cancel one another.

None of the results in the present Letter contains end-point singularities ($x\rightarrow 0$ or $1$), even in the photoproduction limit: the presence of non-zero transverse momenta in $t-$channel allows one to regularize such singularities.

The remaining dependence of the amplitude on the factorization and renormalization scales $\mu_F$ and $\mu_R$
and on the arbitrary parameter $s_0$ is only of 
next-to-next-to-leading logarithmic orders.

Our results were obtained for arbitrary kinematics, in the shockwave approach.
It would be interesting to compare them (in the linear limit for the double dipole contribution) with the result of Ref.~\cite{Ivanov:2004pp}, which was obtained with forward kinematics and for a longitudinally polarized photon, in the usual $k_t-$factorization framework of linear BFKL.
Still, a detailed comparison is not straightforward since 
the distribution of radiative corrections between 
the kernel and the impact factor is different in BK and in BFKL frameworks~\cite{Fadin:2007de,*Fadin:2009gh,*Fadin:2011jg,*Fadin:2012my}. Nontrivial kernel and impact factor transformations are required for such a comparison,
which is left for further studies~\cite{Boussarie:2017}.

\paragraph{Conclusion.}

In this Letter, we have obtained for the first time the complete
NLO impact factor for the $\gamma^{(*)}_{L,T} \to V_L$ transitions
in the shockwave framework. 

The present result, when combined with
solutions to the NLO BK-JIMWLK evolution  and to the leading twist NLO ERBL equation, allows for the very first complete NLO study of exclusive meson production at asymptotic energies with the inclusion of saturation effects.

It paves the way for precision studies of small-$x$ QCD and saturation physics of nucleon and nuclei with a diverse range of phenomenological applications for present and future colliders.   
\\

\acknowledgments

We thank T.~Altinoluk, V.~F.~Fadin, K.~Golec-Biernat and G.~P.~Korchemsky for stimulating discussions.

A.~V.~Grabovsky acknowledges support of president scholarship 171.2015.2, RFBR grants 16-02-00888, 17-02-01187 and 17-52-150024, Dynasty foundation, Metchnikov grant and University Paris Sud. 
R.~Boussarie, A.~V.~Grabovsky, L.~Szymanowski
are grateful to LPT Orsay for hospitality and support while part of the presented work was being done. 
D.~Yu.~Ivanov 
acknowledges support from 
RFBR-15-02-05868 grant, and
thanks the organizers of the 2016 International Summer School of QCD and the GDR 3753 - QCD for support.
This work was partially supported by the ANR PARTONS (ANR-12-MONU-0008-01), the COPIN-IN2P3 Agreement. R.~Boussarie and L.~Szymanowski were supported by grant No 2015/17/B/ST2/01838 of the National Science Center in Poland and by Polish-French
Polonium agreements.
L.~Szymanowski has been supported by the Labex P2IO.

\end{document}